\documentclass{simauth}

\usepackage{amsfonts,amsmath,amssymb}
\usepackage[T1]{fontenc}
\newtheorem{rem}{Remark}
\newcommand{\1}{{\rm 1}\kern-0.25em{\rm I}}
\renewcommand{\P}{{\rm I}\kern-0.22em{\rm P}}
\newcommand{\E}{{\rm I}\kern-0.22em{\rm E}}
\newcommand{\SSS}{{\mathbb S}_n}
\newcommand{\SSSks}{{\mathbb S}_{n,\textrm{ks}}}
\newcommand{\SSSuks}{{\mathbb S}_{n,\textrm{uks}}}
\newcommand{\EE}{{\mathcal E}}
\newcommand{\OO}{{\mathcal O}}

\newcommand{\R}{{\rm I}\kern-0.20em{\rm R}}

\begin{document}
\SIM{1}{6}{00}{28}{00}

\runningheads{Viallon et al.} {How to evaluate the calibration of
a disease risk prediction tool}

\title{How to evaluate the calibration
of a disease risk prediction tool.}

\author{Vivian Viallon\affil{1,2,3}\comma\corrauth, Jacques Bénichou\affil{4}, Françoise
Clavel-Chapelon\affil{1}, Stéphane Ragusa\affil{3}}

\address{\affilnum{1}\ INSERM (Institut de la Santé Et de la Recherche Médicale), ERI 20, Villejuif, F-94805
FRANCE\\\affilnum{2}\ Unité de Biostatistique, Hôpital Cochin, Université Paris-Descartes, Paris, F-75014 FRANCE\\
\affilnum{3}\ Laboratoire de Statistique Théorique et Appliquée
(LSTA), Université Paris VI, Paris, FRANCE\\ \affilnum{4}\ Unité
de Biostatistique, CHU et Faculté de Médecine-Pharmacie de Rouen,
Rouen, FRANCE}

\corraddr{Vivian Viallon, E3N ERI-20, Institut Gustave Roussy, 39,
rue Camille Desmoulins,  Villejuif, F-94805  FRANCE. mail :
vivian.viallon@univ-paris5.fr; phone : + 33 1 58 41 19 60; Fax : +
33 1 58 41 19 61.}



\noreceived{} \norevised{}\noaccepted{}
\begin{abstract}
\noindent To evaluate the calibration of a disease risk prediction
tool, the quantity $E/O$, i.e., the ratio of the expected number
of events to the observed number of events, is generally computed.
However, because of censoring, or more precisely because of
individuals who drop out before the termination of the study, this
quantity is generally unavailable for the complete population
study and an alternative estimate has to be computed. In this
paper, we present and compare four methods to do this. We show
that two of the most commonly used methods generally lead to
biased estimates. Our arguments are first based on some theoretic
considerations. Then, we perform a simulation study to highlight
the magnitude of the previously mentioned biases. As a concluding
example, we evaluate the calibration of an existing predictive
model for breast cancer on the E3N-EPIC cohort. \vskip5pt

\end{abstract}

\section{INTRODUCTION}
Researchers, physicians, as well as the general public, are
focusing increasingly on statistical models designed to predict
the occurrence of a disease. The first corresponding model -- the
Framingham Coronary Risk Prediction Model published in $1976$
\cite{fram} -- was aimed at predicting the individual's risk of
developing heart disease. Modified versions of this primary model
are now widely used by physicians to make decisions on prevention
and treatment strategies. From the late 1980's, researchers
published prediction models for the absolute risk of breast cancer
\cite{roco2}, \cite{gail1}, \cite{roco1},  and some prediction
tools dealing with other types of cancer have begun to appear in
the literature over recent years \cite{harvard1}, \cite{spitz}. In
a workshop held in 2005, Freedman et al. \cite{freed1} already
pointed out the growth of both the number of cancer risk
prediction tools and the need to ensure that they are rigorously
evaluated.\\
Two main criteria, discrimination and calibration, are usually
retained for evaluation. Other criteria may be retained for
particular purposes, see \cite{GailPfeiffer} for some relevant
examples. Discrimination measures the ability to segregate the
individuals into two groups, those who will develop the disease,
and those who will not. It is often evaluated by the concordance
statistic, which is also the area under a receiver operating
characteristic (ROC) curve. Calibration -- our concern here --
measures the ability to predict the number of events in the
population of interest ${\mathbb S}$, usually over a $t_0$-year
period, $t_0>0$: it measures the \emph{goodness-of-fit} of the
model. Calibration is commonly evaluated by comparing the observed
number of events with the number of events expected to occur
within the $t_0$-year period \cite{rock1}, \cite{rock2}. By
summing the estimated $t_0$-year risks over all individuals
belonging to a given representative sample $\SSS$ of the
population ${\mathbb S}$, we get the expected number of cases $E$.
Considering in its turn the number $O$ of cases observed in
${\SSS}$ over the $t_0$-year period, the $E/O$ ratio provides an
estimator of the theoretical quantity $\EE/\OO$ that would be
obtained by evaluating the considered model on the whole
population ${\mathbb S}$, assumed to be infinite (in this
asymptotic setting, $\EE$ and $\OO$ would stand for rates rather
than numbers). A well calibrated model on ${\mathbb S}$ would have
a theoretic $\EE/\OO$ equalling 1. Thus, the $E/O$ ratio is
usually statistically compared to one to definitely assess the
model calibration.\\
However, due to either administrative reasons (the patient was
followed until the end of the study but did not develop the
disease by that date) or the \emph{dropping out} phenomenon
(including both "pure" loss of follow-up and death for reasons
other than the considered disease) data are censored in most
epidemiologic studies. This implies that the $t_0$-year status
regarding the disease is unknown for some individuals, and that
the only available information for these individuals is that they
did not develop the disease after $z$ years of follow-up, with
$0<z<t_0$. In other words, the number of cases which would have
occurred in the population ${\SSS}$ over the $t_0$-year period is
unknown, because of the individuals who dropped out before $t_0$
years of follow-up. To get round this issue, various methods have
been proposed and applied to provide estimators alternative to the
\emph{unobserved} $E/O$ ratio. However, as will be shown later,
most of these methods generally lead to biased estimates. In the
following Section \ref{sec_Methods}, we provide the derivation of
four methods. For each of them we explain its principle as well as
its potential inaccuracy from a theoretical point of view. The
confidence bands associated with each method are also presented.
Then, in Section \ref{simul}, a comparison between the four
methods is performed on simulated data. Finally we compare these
methods on a real sample, the E3N-EPIC cohort, in which we
evaluate one of the Nurses Health Study based breast cancer
prediction tools \cite{roco1} (see Section \ref{example}). These
examples support our assertion that two most commonly used methods
lead to potentially highly biased estimates.

\section{METHODS}\label{sec_Methods}
\subsection{Notations}
Some notations will be of particular interest to describe the
various methods that have been (or can be) used to evaluate
calibration. \vskip3pt

\noindent Let $Y$ be the random variable of interest (in most
cases, $Y$ will stand for the delay between the inclusion in the
study and the occurrence of the considered disease), and $C$ the
censoring variable. The observed variables will be denoted by
$Z=\min(Y,C)$ and $\delta=\1\{Y\leq C\}$, where $\1_{\mathcal S}$
equals 1 if the condition ${\mathcal S}$ is true and 0 otherwise
(i.e., here, $\delta$ equals $1$ if $Y\leq C$, and 0 otherwise).
We fix $t_0>0$ and consider the evaluation of a $t_0$-year risk
prediction tool $P_{t_0}$ on a given population ${\mathbb S}$.
Assume a representative sample $\SSS=\{1,...,n\}\subset {\mathbb
S}$, $n\geq 1$, is at our disposal. In this setting, our aim is to
estimate the theoretical $\EE/\OO$ ratio (relative to $P_{t_0}$ on
${\mathbb S}$) on the sample $\SSS$. For every individual
$i\in\SSS$, denote by $z_i$ his observed time of follow-up, and
$e_i=e_i(t_0)$ his expected risk according to $P_{t_0}$.
Throughout, we will assume that $t_0\leq \max_{i\in\SSS}z_i$.
Further introduce the random variable $O_i=\1\{Y_i\leq t_0\}$
(i.e., $O_i$ equals one if $Y_i\leq t_0$ and 0 otherwise), and set
$o_i$ for the realisation of $O_i$, $i=1,...,n$. \vskip3pt

\noindent We will denote by ${\SSSks}$ the group consisting of
individuals for whom the status regarding the disease after $t_0$
years of follow-up is known. They will be referred to hereafter as
'known $t_0$-status individuals'. This group consists of
\begin{enumerate}
\item individuals who developed the disease before $t_0$ years of
follow-up ($y_i\leq c_i$, $z_i\leq t_0$ and $o_i=1$ for these
individuals);
\item individuals who developed the disease after $t_0$
years of follow-up ($y_i\leq c_i$, $z_i\geq t_0$ and $o_i=0$ for
these individuals);
\item individuals who did not develop the disease and were followed-up at least $t_0$
years ($c_i\leq y_i$, $z_i\geq t_0$ and $o_i=0$ for these
individuals).
\end{enumerate}
Here and elsewhere, $y_i$ [resp. $c_i$] stands for the realisation
of the random variable $Y_i$ [resp. $C_i$]. \vskip3pt

\noindent Similarly, we will denote by ${\SSSuks}$ the group
consisting of individuals for whom the status is unknown: for
these 'unknown $t_0$-status individuals', $c_i\leq y_i$, $z_i\leq
t_0$, so $O_i$ is unobserved and $o_i$ is unknown.\vskip3pt

\noindent Note that the rate of unknown $t_0$-status individuals
increases as $t_0$ increases. Therefore, the size of ${\SSSuks}$
relatively to that of ${\SSSks}$ increases as $t_0$ increases. In
addition, $\SSSks$ as well as $\SSSuks$ are unrepresentative with
respect to the whole population $\SSS$, as generally,
\begin{eqnarray}
 \P(Y_i\leq t_0|i\in\SSSks)&\neq&
\P(Y_i\leq t_0|i\in\SSS)\nonumber \\
\mbox{and} \quad \P(Y_i\leq t_0|i\in\SSSuks)&\neq& \P(Y_i\leq
t_0|i\in\SSS).\label{biais_uks1}
\end{eqnarray}
More precisely, one can see that $\SSSks$ overrepresents cases
with respect to $\SSS$, i.e.,
\begin{equation}\label{biais_uks2}
\P(Y_i\leq t_0|i\in\SSSks)\geq \P(Y_i\leq t_0|i\in\SSS).
\end{equation}
This is the main reason why, to derive inference on ${\mathbb S}$
in the presence of censoring, typical tools (e.g., the
Kaplan-Meier estimate when estimating the unconditional
probability of developing the disease) are required to ensure
unbiased estimates. \vskip3pt

\noindent Before presenting the four methods aimed at evaluating
the calibration of $P_{t_0}$ on ${\mathbb S}$, some additional
notations are needed. Denote by $n_{\textrm{ks}}$ [resp.
$n_{\textrm{uks}}$] the number of individuals belonging to
$\SSSks$ [resp. $\SSSuks$]. Obviously, we have
$n=n_{\textrm{ks}}+n_{\textrm{uks}}$ (since
$\SSSks\cup\SSSuks=\SSS$ and $\SSSks\cap\SSSuks=\emptyset$). The
following quantities will be of particular interest in the sequel.
Introduce
\begin{eqnarray*}
&&E_{\SSS}=\sum_{i\in\SSS}e_i;\quad
E_{\SSSks}=\sum_{i\in\SSSks}e_i;\quad
E_{\SSSuks}=\sum_{i\in\SSSuks}e_i;\\
&&O_{\SSS}=\sum_{i\in\SSS}O_i;\quad
O_{\SSSks}=\sum_{i\in\SSSks}O_i;\quad
O_{\SSSuks}=\sum_{i\in\SSSuks}O_i.
\end{eqnarray*}
Note that the "$O$" terms are random and possibly unobserved (in
particular, $O_{\SSS}$ and $O_{\SSSuks}$ are unobserved) whereas
the "$E$" terms are non-random and known. This is classical in
evaluation studies where inference is made given the sample, which
ensures that $e_i$ is non-random for $i=1,...,n$.\vskip5pt

\noindent Since $O_{\SSS}$ is unobserved, $E_{\SSS}/O_{\SSS}$
cannot be used to estimate the theoretical $\EE/\OO$ ratio. We
present four methods to get round this issue in the following
paragraphs.

\subsection{Method $M_0$}
In some validation studies \cite{rock1},\cite{rock2}, the
evaluation of the calibration is restricted to $\SSSks$, and the
quantity $\EE/\OO$ is estimated by
\begin{equation} \label{est_M0}
\mathcal{R}_{0,n}=E_{\SSSks}/O_{\SSSks}.
\end{equation}
However, in view of (\ref{biais_uks1})$-$(\ref{biais_uks2}), if a
model has to be evaluated on ${\SSS}$, a lot of attention has to
be paid when the validation is performed on ${\SSSks}$: if the
score is well calibrated on ${\mathbb S}$ (and then on $\SSS$),
then the expectation of the $E_{\SSSks}/O_{\SSSks}$ ratio does not
equal 1. In fact, it can even be shown that this expectation is
less than 1, since the known $t_0$-status group ${\SSSks}$
overrepresents cases with respect to $\SSS$ and ${\mathbb S}$ (see
(\ref{biais_uks2}) above and (\ref{biais_M0}) below).

\subsection{Method $M_1$}\label{section_M1}

\noindent Another method, evaluating the calibration on the whole
population ${\SSS}$, can be found in the literature (see, for
instance, \cite{costan1}). The underlying idea is that although
$O_{\SSS}$ is not at our disposal, $O_{1,\SSS}=
\sum_{i\in\SSS}\1\{Y_i\leq \min(z_i,t_0)\}$ is. Then, setting
\begin{equation}\label{E_SSS_M1}
E_{1,\SSS}=\sum_{i\in\SSS}e_i(\min(t_0,z_i)),
\end{equation}
the estimate of $\EE/\OO$ is computed as follows
\begin{equation} \label{est_M1}
\mathcal{R}_{1,n}=E_{1,\SSS}/O_{1,\SSS}=E_{1,\SSS}/O_{\SSSks}.
\end{equation}
Note that $e_i(\min(t_0,z_i))=e_i(t_0)$ only for individuals $i$
who were still disease-free after $t_0$ years. For all other
individuals (i.e., the individuals belonging to $\SSSuks$, plus
the individuals belonging to $\SSSks$ for whom $o_i=1$), we have
$e_i(\min(t_0,z_i))=e_i(z_i)\leq e_i(t_0)$.\\
To see why this method is inappropriate, we present a simple
example. Assume a database of 10,000 individuals followed over a
5-year period (with no dropping-out) is at our disposal. Further
suppose that the risk of the considered disease is uniform over
the 5-year period, such that $\P(Y\leq t)=t/100$, for all $t\leq
5$. Then about 100 cases are likely to be observed each year.
Assume 100 cases are observed each year (giving 500 cases observed
overall) and the evaluation of the prediction tool $P_{t}=t/100$,
for all $t\leq 5$, is under study. All the 9,500 individuals who
remained free from disease after the 5-year period contribute to
5\% in the calculus of $E_{1,\SSS}$. On the other hand, all the
individuals who developed the disease within the first year of
follow-up contribute to (at most) 1\%, those who developed the
disease within the second year of follow-up to 2\%, and so on.
Therefore, according to $M_1$, the number of expected cases is (at
most)
\begin{equation*}
100\times 1\% + 100\times 2\% + 100\times 3\% + 100\times 4\% +
100\times 5\% + 9,500\times 5\%= 490,
\end{equation*}
in such a way that $\mathcal{R}_{1,n}=0.98$! Obviously, the bias
is more severe when the disease prevalence is high.

\subsection{Method $M_2$}
An easy way to correct the aforementioned bias pertaining to $M_1$
exists. In fact, $O_{\SSSks}$ is known and only $O_{\SSSuks}$ is
unknown. Following the idea of $M_1$, however,  $ O_{1,\SSSuks}=
\sum_{i\in\SSSuks}\1\{Y_i\leq \min(z_i,t_0)\} (=
\sum_{i\in\SSSuks}\1\{Y_i\leq z_i\} =0)$, and thus,
$O_{2,\SSS}=O_{\SSSks}+O_{1,\SSSuks}$ are known. Therefore,
setting
\begin{equation}\label{E_SSS_M2}
E_{2,\SSS}=\sum_{i\in\SSSks}e_i(t_0)+\sum_{i\in\SSSuks}e_i(z_i),
\end{equation}
a new estimate of $\EE/\OO$ is given by
\begin{equation} \label{est_M2}
\mathcal{R}_{2,n}=E_{2,\SSS}/O_{2,\SSS}=E_{2,\SSS}/O_{\SSSks}.
\end{equation}
Concretely, in (\ref{E_SSS_M2}), the individuals who developed the
disease before $t_0$ years of follow-up contribute to $e_i(t_0)$
while they contribute to $e_i(z_i)$ in (\ref{E_SSS_M1}) (keep in
mind that $e_i(z_i)\leq e_i(t_0)$ because $z_i\leq t_0$ for such
individuals). Comparing the estimates provided by $M_1$ and $M_2$,
it is easily derived that
\begin{equation}\label{biais_M1}
\mathcal{C}_1(t_0)=\frac{\mathcal{R}_{2,n}}{\mathcal{R}_{1,n}}=1 +
\frac{\sum_{i\in\SSSks}\delta_i
\{e_i(t_0)-e_i(z_i)\}}{O_{\SSSks}}\geq 1.
\end{equation}
Note that, to our knowledge, $M_2$ has never been used so far,
although it provides a simple and practical way to improve $M_1$.
However, it is not clear whether $\mathcal{R}_{2,n}$ is unbiased
or not: the explicit expression of the expectation of
$\mathcal{R}_{2,n}$ (or $1/\mathcal{R}_{2,n}$) can not be easily
derived. Moreover, there exists a drawback common to both $M_1$
and $M_2$: using either method, the evaluation of a crude
$t_0$-year risk score can not be performed. In fact, some of the
$e_i$'s involved in the calculation of $E_{1,\SSS}$ and
$E_{2,\SSS}$ are attached to a $t_0$-year period, while others are
attached to a $z_i$-year period. The main problem arises when
calibration is adjusted for percentiles of predicted risk (which
is quite common in evaluation studies), and is due to the fact
that the $e_i$'s are not comparable. In this adjusted setting, the
estimation of the $e_i$'s distribution, and then the derivation of
their percentiles, becomes hazardous. Similar problems also arise
when calibration is adjusted for risk factors (such as age at
inclusion or personal history of the disease). Therefore, $M_2$
should not be used when adjusted calibration has to be evaluated.

\subsection{The method $M_3$}
Keep in mind that in the absence of dropping-out, the quantity
$E_{\SSS}/O_{\SSS}$ provides a suitable estimate of $\EE/\OO$. The
problem in the presence of dropping-out arises from the fact that
$O_{\SSS}$ is unknown. However, a natural candidate to replace
$O_{\SSS}$ is defined as follows,
\begin{equation}
\widehat{O}_{\SSS}=n K_n(t_0),\label{approx1}
\end{equation}
where $K_n(t_0)$ is the Kaplan-Meier estimate of $\P(Y\leq t_0)$
on ${\SSS}$. Using this \cite{ferrario}, an estimate of $\EE/\OO$
can be given by
\begin{equation}
{\mathcal R}_{3,n}= \frac{E_{\SSS}}{\widehat{O}_{\SSS}}\cdot
\label{est_M3}
\end{equation} Note that since $K_n(t_0)\rightarrow \P(Y\leq t_0)$ almost
surely as $n\rightarrow\infty$, for any $t_0\leq
\max_{i\in\SSS}z_i$ \cite{StuteWang}, it is easily derived that
${\mathcal R}_{3,n}$ is asymptotically unbiased. \vskip7pt

Through this theoretical description of the various methods, we
showed that ${\mathcal R}_{0,n}$ and ${\mathcal R}_{1,n}$ provide
biased estimates. Moreover, the asymptotic unbiasedness was
established for ${\mathcal R}_{3,n}$  but not for ${\mathcal
R}_{2,n}$, suggesting that ${\mathcal R}_{3,n}$ is the most
reliable estimate of the $\EE/\OO$ ratio. Moreover, ${\mathcal
R}_{3,n}$ is intuitively the most appealing estimator because it
takes into account all the information available after $t_0$ years
of follow-up. These statements will be confirmed by the simulation
studies performed in Section \ref{simul}.

\subsection{Some complements}
Some additional properties of the various methods merit
presentation.

\subsubsection{The inadmissibility of $M_0$}
Comparing $\mathcal{R}_{0,n}$ and $\mathcal{R}_{3,n}$ gives
insight into the magnitude of the bias pertaining to $M_0$. Under
the assumption of independence between the vector of covariates
and the censoring variable, it can be shown that
\begin{equation}\label{biais_M0}
{\mathcal C}_0(t_0) = \frac{\mathcal{R}_{3,n}}{\mathcal{R}_{0,n}}
= \frac{E_{\SSS}/\widehat{O}_{\SSS}} {E_{\SSSks} /
O_{\SSSks}}\approx
\frac{F_{n_{\textrm{ks}}}(t_0)}{K_n(t_0)}=\widetilde{{\mathcal
C}}_0(t_0),
\end{equation}
where $F_{n_{\textrm{ks}}}(t_0)$ is the empirical distribution
function on $\SSSks$, i.e., the standard estimate of the
probability of developing the disease on ${\SSSks}$. See Appendix
for the proof of (\ref{biais_M0}).\vskip3pt

\noindent Since $K_n(t_0) \leq F_{n_{\textrm{ks}}}(t_0)$, almost
surely for $n$ large enough, we have ${\mathcal C}_0(t_0) \geq 1$.
For instance, set $\mathcal{Z}=\max_{i\in\SSS} z_i$, and select
$t_0=\mathcal{Z}$. In this particular example, all the cases
belong to ${\SSSks}$; non-cases do not. Thus,
$F_{n_{\textrm{ks}}}(\mathcal{Z}) = 1$, while $K_n(\mathcal{Z})\ll
1$ (typically, $K_n(\mathcal{Z})$ does not exceed 0.2), and
$\mathcal{C}_0 (\mathcal{Z})$ becomes high. Even if this case is
somewhat extreme, it highlights the inadmissibility of $M_0$,
which is however among the most widely used of methods.\vskip5pt

\subsubsection{Confidence intervals}
In order to conclude whether a given prediction tool is well
calibrated on $\mathbb{S}$ or not, confidence intervals are
generally needed. When the estimation of $\EE/\OO$ is based on
$M_0$, $M_1$ or $M_2$, such intervals can be calculated using the
Poisson variance  for the logarithm of the observed number of
cases \cite{rock2}. Namely, for $j=0,1,2,$
\begin{equation}\label{IC}
\mbox{CI}_{j,n,95\%}(\EE/\OO)= \Big[{\mathcal
R}_{j,n}\exp\Big(\pm1.96 \sqrt{1/O_{\SSSks}}\Big)\Big].
\end{equation}
Note that, since $M_0$ and $M_1$ lead to biased estimates of the
quantity $\EE/\OO$, the above formula may only be correct for
$j=2$ (if, eventually, $\mathcal{R}_{2,n}$turns out to be
unbiased).\\
On the other hand, in the case of $M_3$, a log-transformation can
be coupled with the delta-method, giving
\begin{equation*}
\textrm{Var} \Big[\log \Big(\frac{E_{\SSS}}
{\widehat{O}_{\SSS}}\Big)\Big]=\frac{\sigma^2_{n,t_0}}{K^2_n(t_0)},
\end{equation*}
where $\sigma^2_{n,t_0}$ is the Greenwood variance
\cite{kabfleischprentice} of the Kaplan-Meier estimate evaluated
at $t_0$. The corresponding confidence interval is given by
\begin{eqnarray}
\mbox{CI}_{3,n,95\%}(\EE/\OO) = \Big[{\mathcal
R}_{3,n}\exp\Big(\pm1.96 \frac{\sigma_{n,t_0}}{K_n(t_0)}
\Big)\Big] \cdot \label{IC1}
\end{eqnarray}

\section{SIMULATION STUDY}\label{simul}

A simulation study was performed to check that $\mathcal{R}_{2,n}$
and $\mathcal{R}_{3,n}$ were better estimates of the $\EE/\OO$
ratio than $\mathcal{R}_{1,n}$ and $\mathcal{R}_{0,n}$, and to
compare $\mathcal{R}_{2,n}$ and
$\mathcal{R}_{3,n}$. \\
We considered the case where $Y\rightsquigarrow
\mathcal{U}(0,\lambda)$, for a given $\lambda>t_0$, i.e., $Y$ was
uniformly distributed on the interval $[0,\lambda]$. This ensured
that $\P(Y\leq t)=t /\lambda$, for all $0\leq t\leq \lambda$. Note
that the higher the rate $1/\lambda$, the higher the ``prevalence
of the disease'', and therefore, the higher the bias of ${\mathcal
R}_{1,n}$ is expected to be (see Section \ref{section_M1}). For
the censoring variable, we chose $C\rightsquigarrow
\mathcal{U}(0,\omega_\lambda)$, for a given $\omega_\lambda>0$. To
allow the rate of unknown $t_0$-status individuals to vary, we
selected various values of $\omega_\lambda$, depending on the rate
$1/\lambda$. We also considered the case with no censure (and
therefore with no unknown $t_0$-status individuals) to check that
$M_0$,
$M_2$ and $M_3$ provided the same estimates in this case.\\
Given $n\geq 1$, samples $(Y_1,...,Y_n)$ and, if appropriate,
$(C_1,...,C_n)$ were simulated. From these samples, we generated
the ``observed'' sample $((Z_1,\delta_1),...(Z_n,\delta_n))$,
where, as usual, $Z_i=\min(Y_i, C_i)$ and $\delta_i=\1\{Y_i\leq
C_i\}$. The population $\SSS=\{1,...,n\}$ could then be split into
$\SSSks$ and $\SSSuks$, making the calculation of $O_{\SSSks}$ and
$O_{\SSSuks}$ possible. Moreover, the Kaplan-Meier estimate could
be calculated on our samples, enabling us to compute $\widehat
O_{\SSS}$. Finally,  the terms $E_{\SSS}$, $E_{\SSSks}$,
$E_{1,\SSS}$ and $E_{2,\SSS}$, and then $\mathcal{R}_{0,n}$,
$\mathcal{R}_{1,n}$, $\mathcal{R}_{2,n}$ and $\mathcal{R}_{3,n}$,
were computed using the formula $e_i(t_0)=t_0 /\lambda$ and
$e_i(z_i)=z_i /\lambda$, and the corresponding confidence
intervals were constructed making use equations (\ref{IC}) and
(\ref{IC1}). Note that, given the way the expected number of cases
was calculated, the underlying prediction tool should be well
calibrated, and $\mathcal{R}_{j,n}$ should be close to one if the
method $M_j$, $j=0,1,2,3$, provided unbiased estimates of the
$\EE/\OO$ ratio.\vskip5pt

\noindent In every example, we selected $n=20,000$ and $t_0=10$.
We repeated the procedure described above 1,000 times, computing
$(i)$ the mean for each of the $\mathcal{R}_{j,n}$, $j=0,1,2,3$,
$(ii)$ the mean width of the corresponding confidence interval and
$(iii)$ the proportion of confidence intervals including the value
1 (which is an estimate of the covering probability of the
confidence interval). \vskip5pt

\noindent We selected $\lambda=100$, $\lambda=200$ and
$\lambda=400$, and in each case, we selected three values of
$\omega_\lambda$ such that there was 5\%, 10\% and 20\% of unknown
10-year status individuals (plus the case with no censure at all):
this resulted in $3\times4=12$ simulation designs. The results are
presented in Table \ref{table_simul1}. \vskip3pt

\noindent First consider the mean of the point estimates obtained
for each method in each simulation design. We observed that the
estimates $\mathcal{R}_{0,n}$, $\mathcal{R}_{2,n}$ and
$\mathcal{R}_{3,n}$ were identical in the uncensored cases,
corresponding to the cases where the rate of unknown $t_0$-status
individuals was null. In addition, we observed that
$\mathcal{R}_{1,n} < 1$ in every case, and that the bias magnitude
depended upon the "prevalence" $1/\lambda$, independently of the
rate of unknown $t_0$-status individuals. Finally, the error made
when using $\mathcal{R}_{0,n}$ was all the higher as this rate
increased (as expected again). All these observations confirmed
the assertions presented in Section \ref{sec_Methods}. The
correcting terms presented in Table \ref{table_correc} made these
observations even clearer and supported the approximation stated
in (\ref{biais_M0}). Furthermore, the estimates
$\mathcal{R}_{2,n}$ and $\mathcal{R}_{3,n}$ gave very similar
values, which were close to the true value 1. These first results
confirmed the fact that $\mathcal{R}_{2,n}$ and
$\mathcal{R}_{3,n}$ were better estimates of the $\EE/\OO$ ratio
than $\mathcal{R}_{0,n}$ and $\mathcal{R}_{1,n}$, and then that
the use of the latter two estimators should be avoided. \vskip3pt

\noindent Considering in more detail $\mathcal{R}_{2,n}$ and
$\mathcal{R}_{3,n}$, we saw that the means of the
$\mathcal{R}_{3,n}$'s were slightly closer to 1 that those of the
$\mathcal{R}_{2,n}$'s. Moreover, by comparing the width and the
covering probability of the corresponding confidence interval,
$\mathcal{R}_{3,n}$ appeared to be more precise than
$\mathcal{R}_{2,n}$, with narrower but still more accurate
confidence intervals. Therefore, from this simple simulation
study, the estimate $\mathcal{R}_{3,n}$ turned out to be the most
advisable one. \vskip3pt

\noindent Note that the precision of $\mathcal{R}_{3,n}$ (as well
as that of $\mathcal{R}_{2,n}$) was closely related to the
prevalence $1/\lambda$: the higher the prevalence, the more
precise the estimates.

\section{CASE STUDY : THE EVALUATION OF AN EXISTING BREAST CANCER PREDICTION TOOL ON THE
E3N COHORT}\label{example}

E3N (Etude Epidémiologique des femmes de l'Education Nationale) is
the French component of the EPIC (European Prospective
Investigation into Cancer and nutrition) prospective study and has
been thoroughly described elsewhere \cite{agnes1}. All
participants are women belonging to the Mutuelle Générale de
l'Education Nationale (MGEN), a health insurance scheme primarily
covering teachers, teacher's spouses, and employees of the
National Education System. Since June 1990, after having given
informed consent, 98,995 women have been asked at approximately
24-month intervals to complete self-administered questionnaires,
which include a variety of lifestyle characteristics. After the
exclusion of the prevalent cases of cancer (n=6,999) and women who
had never menstruated ($n=28$), the cohort includes 91,968
observations (with 3,467 cases of invasive breast cancer).
\vskip5pt

\noindent Rosner and Colditz models proposed two breast cancer
risk prediction models according to which incidence of breast
cancer at age $a$ ($I_a$) is proportional to the number of breast
cell divisions accumulated throughout life up to age $a$
\cite{roco1}, \cite{roco2}. The rate of breast cancer cell
division at age $a'$ is supposed to be dependent on risk factors
that are relevant at age $a'$. Rosner and Colditz thus expressed
the log incidence rate of breast cancer as a linear function of
the cumulative effect of individual breast cancer risk factors. In
a first attempt \cite{roco1}, in addition to age ($a$), only
reproductive factors were considered, namely, age at menarche
($a_0$), menopausal status ($m$), age at menopause ($a_m$), parity
$s$, age at first birth $(a_1)$, and a variable $b$, called
\emph{birth index} and defined as $ b=\sum_{i=1}^s
(a^\star-a_i)b_{i,a}$, where $a_i$ is the age at $i$th birth,
$a^\star=\min(a, a_m)$ and $b_{i,a}=1$ if parity is greater than
$i$ at age $a$, $0$ otherwise. Defining $b_1$ as 1 if $s\geq 1$, 0
otherwise, the $RCM$ was specified as
\begin{eqnarray}
\log I_a &=& \alpha + \beta_0a_0 + \beta_1(a^\star-a_0) +
\beta_2(a-a_m)m \nonumber\\
&&\!\!\!\!\!\!+ \beta_3(a_1-a_0)b_1 + \beta_4b +
\beta_5b(a-a_m)m.\label{modele_RCM}
\end{eqnarray}
The values of the parameters $\alpha, \beta_1,..., \beta_5$
estimated in \cite{roco1} are recalled for convenience in Table
\ref{table_coeff}. \vskip3pt

\noindent Rosner and Colditz later developed a model including
more factors \cite{roco2}; an evaluation study of the two versions
can be found in Rockhill et al. \cite{rock2}. We chose to evaluate
the first version ($RCM$) as some of the variables involved in the
extended one were not available in the E3N study (BMI at menarche,
for instance). Moreover, our aim was to measure the respective
performances of the methods presented in Section \ref{sec_Methods}
rather than to evaluate the best published model. \vskip5pt

\noindent To compute the $t$-year risk (where $t$ can take the
value $t_0$ or $z_i$ depending on the method to be used to
evaluate the $RCM$), we proceeded as in Rockhill et al.'s
evaluation study \cite{rock2}: once we obtained, from
(\ref{modele_RCM}), the log incidence rates for each year for each
woman, we exponentiated each one to get an incidence rate $r_j$,
$j=1,...,t$, for each year during the $t$-year period; then, the
$t$-year risk was computed as $1 - \exp(-[r_1+\cdots+r_{t}])$. We
chose $t_0=10$ years. In addition, we performed the evaluation on
three groups:

\begin{itemize}
\item the whole sample ($n=91,968$);
\item Postmeno. Group 1, comprised of women who were at inclusion ($n=36,603$);
\item Postmeno. 2, comprised of Postmeno. Group 1 plus the women who
went through the menopause during the study (these women entered
this group at the time of their menopause) ($n=82,402$).
\end{itemize}

\noindent For the whole sample ($n=91,968$), the rate of unknown
$t_0$-status individuals was 12\% ($n_{\textrm{ks}}=80,883$), and
the Kaplan-Meier estimate of the unconditional risk of disease was
3.21\%. For Postmeno. Group 1 ($n=36,603$), the rate of unknown
$t_0$-status individuals was 12.5\% ($n_{\textrm{ks}}=32,027$),
and the Kaplan-Meier estimate of the unconditional risk of disease
was 3.39\%. For Postmeno. Group 2 ($n=82,402$), the rate of
unknown $t_0$-status individuals was 51.5\%
($n_{\textrm{ks}}=39,931$), and the Kaplan-Meier estimate of the
unconditional risk of disease was 3.60\%. The results presented in
Table \ref{res_M0_tot} were consistent with our previous
explanations. The estimates $\mathcal{R}_{2,n}$ and
$\mathcal{R}_{3,n}$ gave similar results, whereas
$\mathcal{R}_{1,n}$ and especially $\mathcal{R}_{0,n}$ were
slightly different and conceivably biased. Moreover,
$\mathcal{R}_{3,n}$ was more precise than $\mathcal{R}_{2,n}$.
Note that the bias magnitude of $\mathcal{R}_{0,n}$ was of the
same order as that expected in view of the results of the
simulation study. In fact, for the whole sample and Postmeno.
Group 2, the rate of unknown $t_0$-status individuals was about 12
\%. In our simulation study, we observed that
$\widetilde{{\mathcal C}}_0(t_0)\simeq 1.055$ for 10 \% of unknown
$t_0$-status individuals. Here, we had ${\mathcal C}_0(t_0)=
1.065$ and ${\mathcal C}_0(t_0)= 1.074$ on the whole sample and
Postmeno. Group 1 respectively (on Postmeno. Group 2, we had
${\mathcal C}_0(t_0)= 1.43$, but this feature could not be
compared with our simulated results, since the rate of unknown
$t_0$-status individuals reached 51.5\% for this group). However,
the bias magnitude of $\mathcal{R}_{1,n}$ was slightly less
important than what could have been expected from our simulation
study. Indeed, the prevalence of the disease was about $1/300$ per
year (around 1/30 over 10 years), and we calculated ${\mathcal
C}_0(t_0) \simeq 1.01$, while 1.02 was expected. This highlights
the fact that the distribution of $Y$ plays an important role with
respect to the bias of $\mathcal{R}_{1,n}$. In fact, this bias is
larger for a uniform distribution than, for instance, an
exponential one, where cases are likely to occur later (and in
which case, the terms $e_i(z_i)$ are likely to be closer to
$e_i(t_0)$). \vskip3pt

\noindent Note that the $RCM$ appeared to slightly underestimate
the breast cancer risk in the whole E3N population. This
underestimation was wider for the postmenopausal groups,
especially on Postmeno. Group 2. The main reason might be that
$RCM$ does not take hormone replacement therapy (HRT) use into
account. HRT is known to increase the risk of cancer
\cite{agnes2}, \cite{agnes1}. Moreover, the use of HRT is more and
more frequent in the E3N population as well as in the general
population: this means that, overall, the use of HRT is more
frequent in Postmeno. Group 2 than in Postmeno. Group 1.
Therefore, this could explain (at least partly) the wider
underestimation observed on Postmeno. Group 2.

\section{DISCUSSION}
We have presented and compared four methods aimed at evaluating
the calibration of disease risk prediction tools. It was shown
that the estimates $\mathcal{R}_{3,n}$ and $\mathcal{R}_{2,n}$
should be preferred to $\mathcal{R}_{0,n}$ and
$\mathcal{R}_{1,n}$, the latter two being biased in most
situations. The estimator $\mathcal{R}_{3,n}$ appeared to be more
precise than $\mathcal{R}_{2,n}$ on simulated data. In addition,
the unbiasedness of $\mathcal{R}_{2,n}$ was not theoretically
established here, and its applicability was shown to be limited
(in particular, it should not be used when calibration has to be
adjusted for percentiles of predicted risks). \vskip5pt

\noindent Some other more sophisticated criteria (such as Hosmer
and Lomeshow \cite{hosmer} \emph{goodness-of-fit} statistics) may
also be retained to evaluate the calibration. Here, we focused on
the so-called $E/O$ ratio, but the problems arising in this simple
case of course still arise when more sophisticated criteria are
used, and we recommend the use of the Kaplan-Meier estimate to
estimate the "O" terms involved in the Hosmer-Lomeshow statistic.
If this is done, however, we also recommend either checking the
$\chi^2$ distribution of the resulting statistic or using
bootstrap techniques to derive the associated p-value.\vskip5pt

\noindent The problem of individuals who dropped out before $t_0$
years of follow-up still arises when evaluating the discrimination
of a $t_0$-year risk score. It has been shown that the concordance
statistic is biased when estimated only on the known $t_0$-status
group, and an unbiased estimate has been proposed when the
underlying model is a Cox proportional hazard model with time
under study as the time scale \cite{Gonen}. In other cases, no
unbiased estimates have ever been proposed. An alternate approach
is to compute the \emph{Observed Relative Risk} ($ORR$). To do
this, individuals have to be sorted by predicted $t_0$-year risks.
Then, the $ORR$ is simply the ratio of the number of observed
cases in the top decile (or quintile) of predicted $t_0$-year
risks to the number of observed cases in the bottom decile (or
quintile). Obviously, since observed numbers of cases are
generally not at the statistician's disposal, Kaplan-Meier
estimates (and bootstrap confidence intervals) are required in
this setting too. \vskip5pt

\noindent As a conclusion, we strongly recommend the use of
$\mathcal{R}_{3,n}$ as an estimate of the $\EE/\OO$, even if it is
not the most commonly used estimate in the evaluation of
calibration literature (especially in the breast cancer field).

\section{APPENDIX}

\subsection{Proof of (\ref{biais_M0})}
Our aim is first to prove (\ref{biais_M0}), which is recalled in
(\ref{dern}) below for convenience,
\begin{equation}\label{dern}
{\mathcal C}_1(t_0) \approx
\frac{F_{n_{\textrm{ks}}}(t_0)}{K_n(t_0)}.
\end{equation}
First note that
\begin{equation}
\frac{E_{\SSS}} {O_{\SSS}}= \frac{ (E_{{\SSSuks}} + E_{{\SSSks}})
}{O_{{\SSSks}}} \frac{O_{{\SSSks}}}{O_{\SSS}} =
\frac{E_{{\SSSks}}} {O_{{\SSSks}}}\bigg(1 + \frac{E_{{\SSSuks}}}
{E_{{\SSSks}}}\bigg) \frac {O_{{\SSSks}}} {O_{\SSS}}, \nonumber
\end{equation}
where $O_{{\SSSuks}}$ (resp. $E_{{\SSSuks}}$) is the observed
(resp. expected) number of cases on ${\SSSuks}$. \\
Keeping in mind that $O_{{\SSSks}}=n_{\textrm{ks}}
F_{n_{\textrm{ks}}}(t_0)$, $\widehat O_{\SSS} = n K_n(t_0)$) and
$n=n_{\textrm{ks}}+n_{\textrm{uks}}$, it is straightforward that
\begin{eqnarray}
\frac{E_{{\mathbb{S}}}} {\widehat O_{{\mathbb{S}},n}} &=&
\frac{E_{{\SSSks}}} {O_{{\SSSks}}}\bigg(\frac{1 +
{E_{{\SSSuks}}}/E_{{\SSSks}}}{1+n_{\textrm{uks}}/n_{\textrm{ks}}}\bigg)
\frac {F_{n_{\textrm{ks}}}(t_0)} {K_n(t_0)}\cdot \nonumber
\end{eqnarray}
Next, introduce the following assumption:
\begin{tabbing}
$(H)\;\;$ \= The censoring process is independent from the
covariates.
\end{tabbing}
\begin{rem}\label{rema}
The condition $(H)$ ensures that the distribution of the
covariates is the same on ${\SSSuks}$ and ${\SSSks}$ (and then on
${\SSS}$).
\end{rem}
Under $(H)$, with $\hat{e}_n=E_{\SSS}/n$,
\begin{equation}
\frac{E_{{\SSSuks}}}{E_{{\SSSks}}}\simeq \frac{\hat{e}_n
n_{\textrm{uks}}}{\hat{e}_n
n_{\textrm{ks}}}=\frac{n_{\textrm{uks}}}{n_{\textrm{ks}}},
\nonumber
\end{equation}
in such a way that
\begin{equation}
\frac{E_{\SSS}} {\widehat O_{\SSS}} \simeq \frac{E_{{\SSSks}}}
{O_{{\SSSks}}}\frac {F_{n_{\textrm{ks}}}(t_0)} {K_n(t_0)}\cdot
\end{equation}

\begin{table}
\begin{footnotesize}
\caption{Results of the simulation studies. The mean of the
estimate of the $\EE/\OO$ ratio, the mean of the width of the
corresponding confidence interval and the proportion of these
intervals including the value 1 are given respectively for each of
the four methods in each of the 12 simulation designs. Means were
obtained from 1,000 independent samples.}\label{table_simul1}
\begin{center}
\begin{tabular}{l r r @{.} l r @{.} l r @{.} l r @{.} l}
\toprule Rate of & Observed &\multicolumn{2}{c}{Method} &
\multicolumn{2}{c}{Method} & \multicolumn{2}{c}{Method} &
\multicolumn{2}{c}{Method} \\
UKSI$^\ddag$ & Cases & \multicolumn{2}{c}{$M_0$} &
\multicolumn{2}{c}{$M_1$} & \multicolumn{2}{c}{$M_2$} &
\multicolumn{2}{c}{$M_3$}\\
\midrule \multicolumn{3}{l}{$Case\ 1 : \lambda=100$}\\[0.2cm]
$0$&   2,000 & 1&000 0.088 0.967 & 0&950 0.083 0.374 & 1&000 0.088 0.967 & 1&000 0.083 0.957\\
$5\%$& 1,947 & 0&976 0.087 0.809 & 0&951 0.085 0.406 & 1&001 0.089 0.955 & 1&001 0.084 0.946\\
$10\%$&1,895 & 0&950 0.086 0.391 & 0&951 0.086 0.410 & 1&002 0.090 0.967 & 1&001 0.086 0.961\\
$20\%$&1,787 & 0&893 0.083 0.003 & 0&953 0.088 0.462 & 1&004 0.093 0.963 & 1&001 0.088 0.951\\[0.2cm]
 \multicolumn{3}{l}{$Case\ 2 : \lambda=200$}\\[0.2cm]
$0$&  1,001  &1&000 0.124 0.954 & 0&975 0.121 0.876  & 1&000 0.124 0.954 & 1&000 0.121 0.949\\
$5\%$& 973 & 0&976 0.123 0.891 & 0&978 0.123 0.891 & 1&003 0.126 0.960 & 1&002 0.123 0.953\\
$10\%$&948 & 0&949 0.121 0.620 & 0&976 0.124 0.898 & 1&002 0.128 0.958 & 1&001 0.124 0.956\\
$20\%$&896 & 0&890 0.117 0.066 & 0&977 0.128 0.887 & 1&003 0.132 0.959 & 1&001 0.128 0.955\\[0.2cm]
\multicolumn{3}{l}{$Case\ 3 : \lambda=400$}\\[0.2cm]
$0$&   500 & 1&002 0.176 0.950 & 0&990 0.174 0.931 & 1&002 0.176 0.950 & 1&002 0.174 0.950\\
$5\%$& 488 & 0&976 0.174 0.907 & 0&989 0.176 0.939 & 1&002 0.178 0.964 & 1&002 0.181 0.960\\
$10\%$&475 & 0&948 0.171 0.783 & 0&989 0.178 0.942 & 1&001 0.181 0.968 & 1&001 0.178 0.964\\
$20\%$&448 & 0&893 0.166 0.313 & 0&992 0.184 0.965 & 1&005 0.187 0.968 & 1&004 0.185 0.966\\
\bottomrule
\end{tabular}
\end{center}
$^\ddag$ Unknown $t_0$-status individuals.
\end{footnotesize}
\end{table}

\begin{table}
\begin{footnotesize}
\caption{Results of the simulation studies showing the mean of the
correction terms. Means were obtained from 1,000 independent
samples.}\label{table_correc}
\begin{center}
\begin{tabular}{l c c}
\toprule Rate of & Correction term &  Correction term \\
UKSI$^\ddag$ & $\widetilde{{\mathcal C}}_0(t_0)^\dag$   & ${\mathcal C}_1(t_0)^*$  \\
\midrule \multicolumn{1}{l}{$Case\ 1 : \lambda=100$}\\[0.2cm]
$0$&  1.000 &  1.053 \\
$5\%$& 1.025 & 1.053\\
$10\%$& 1.053& 1.054\\
$20\%$& 1.120& 1.055\\[0.2cm]
 \multicolumn{1}{l}{$Case\ 2 : \lambda=200$}\\[0.2cm]
$0$&  1.000 &  1.026 \\
$5\%$& 1.026 & 1.026\\
$10\%$& 1.055& 1.026\\
$20\%$& 1.124& 1.027\\[0.2cm]
\multicolumn{1}{l}{$Case\ 3 : \lambda=400$}\\[0.2cm]
$0$&  1.000 &  1.013 \\
$5\%$& 1.026 & 1.013\\
$10\%$& 1.055& 1.013\\
$20\%$& 1.125& 1.013\\[0.2cm]
\bottomrule
\end{tabular}
\end{center}
$^\ddag$ Unknown $t_0$-status individuals.\\
$^\dag \widetilde{{\mathcal C}}_0(t_0) = F_{n_{\textrm{ks}}}(t_0)/K_n(t_0)$. \\
$^*{\mathcal C}_1(t_0) = \mathcal{R}_{2,n}/\mathcal{R}_{1,n}$.\\
\end{footnotesize}
\end{table}

\begin{table}
\begin{footnotesize}
\caption{Coefficients of the first Rosner and Colditz model
($RCM$).}\label{table_coeff}
\begin{center}
\begin{tabular}{l r @{.} l l}
\toprule Parameter & \multicolumn{2}{c}{Regression coefficient} &
SE* \\
\midrule
$\alpha$ \,\,(intercept) & -9&687 & 0.265 \\
$\beta_0$ (age at menarche) & 0&048 & 0.016 \\
$\beta_1$ (min[age, age at menopause] $-$ age at menarche)& 0&081 & 0.004 \\
$\beta_2$ (age $-$ age at menopause), for menopausal women & 0&050 & 0.005 \\
$\beta_3$ (age at first birth $-$ age at menarche) & 0&013 & 0.004 \\
$\beta_4$ (birth index) & -0&0036 & 0.0009 \\
$\beta_5$ (birth index $\times$ [age $-$ age at menopause]), for menopausal women & -0&00020 & 0.00012 \\
\bottomrule
\end{tabular}\\
*SE: standard error.
\end{center}
\end{footnotesize}
\end{table}

\begin{table}
\begin{footnotesize}
\caption{Evaluation of the calibration of the Rosner and Colditz
10-year risk of breast cancer prediction tool. Results from the
E3N cohort.}\label{res_M0_tot}
\begin{center}
\begin{tabular}{l c c c c c c}
Population & Rate of & Observed &Method & Method & Method &
Method \\
for validation &UKSI$^\ddag$ & Cases & $M_0$ & $M_1$ & $M_2$ &
$M_3$\\& & & [CI*] & [CI*] & [CI*] &[CI*] \\
\midrule
Whole sample & 12.1\% & 2,765 & 0.889  & 0.932  & 0.940  & 0.947  \\
&&&[0.839-0.941]&[0.880-0.987] & [0.887-0.996]
 &[0.912-0.982] \\
 Postmeno. Group 1$^\dag$ &12.5\% & 1,160 & 0.635  & 0.672  & 0.678 & 0.682 \\
&&&[0.600-0.673]&[0.634-0.711] & [0.640-0.718]
 &[0.644-0.721]\\
Postmeno. Group 2$^\natural$ & 51.5\% &2,115 &0.417  & 0.591 & 0.597 & 0.595 \\
&&&[0.394-0.442]& [0.558-0.626] & [0.564-0.633]
 &[0.569-0.620] \\
\bottomrule
\end{tabular}
\end{center}
$*$ CI : Confidence intervals.\\
$^\ddag$ Unknown $t_0$-status individuals.\\
$^\dag $Postmenopausal women at inclusion.\\
$^\natural$Postmenopausal women during follow-up.
\end{footnotesize}
\end{table}

\end{document}